\documentclass[12pt]{article}

\usepackage{amssymb}
\usepackage{color}
\usepackage{epsfig,amssymb,amsfonts,amsmath,graphicx,dsfont,cite,xfrac}
\usepackage{authblk}
\usepackage{subfigure}
\definecolor{mygray}{gray}{0.5}

\usepackage{cite}
\usepackage[colorlinks=true,linkcolor=blue,citecolor=red]{hyperref}

\newcommand{\be}{\begin{equation}}
\newcommand{\ee}{\end{equation}}
\newcommand{\bea}{\begin{eqnarray}}
\newcommand{\eea}{\end{eqnarray}}

\title{Exactly Solvable Time-Dependent Oscillator-Like Potentials Generated by Darboux Transformations}

\author[${}$]{Kevin Zelaya}
\author[${}$]{Oscar Rosas-Ortiz}

\affil[${}$]{\footnotesize Physics Department, Cinvestav, AP 14-740, 07000
M\'exico City, Mexico}

\date{}
\begin{document}

\maketitle

\begin{abstract}
The stationary Schr\"odinger equation of the harmonic oscillator is deformed by a Darboux transformation to construct time-dependent potentials with the oscillator profile. The Darboux (supersymmetric or factorization) method is usually developed in the spatial variables of the Schr\"odinger equation. Here we follow a variation introduced by Bagrov, Samsonov and Shekoyan to include the time-variable as a parameter of the transformation.
\end{abstract}


\section{Introduction}

It is well known that the Darboux transformations leave key geometric properties of certain classes of surfaces unchanged \cite{Rog02}. Such a property finds a diversity of applications in physics because the sets of solutions of the differential equations that represent dynamical laws can be modeled in geometric form. One of the branches of physics which uses the Darboux transformation in exhaustive form is the soliton theory, where nonlinear superposition principles hold. Unexpectedly, the models introduced in the eighties of the previous century to study (in the same picture) bosons and fermions were also associated with the Darboux transformation \cite{Mie04}. The term {\em supersymmetric quantum mechanics} came to denote the simplest case of such models and gave rise to a new branch of quantum physics which has grown stronger over the years \cite{Mie04,Bag01,Coo01,Gan11}. In this context, most of the works dealing with the supersymmetric construction of exactly solvable potentials use the Darboux transformation of a given potential in the spatial variable of the Schr\"odinger equation. A notable exception is offered by the papers \cite{Bag95,Bag96}, where the authors introduce a variation of the Darboux transformation that includes the time-variable as a parameter. Using such a method, one can construct time-dependent potentials which are exactly solvable.

In this paper we report a new family of exactly solvable time-dependent potentials that correspond to the Darboux-deformations of the harmonic oscillator in the approach of Refs.~\cite{Bag95,Bag96}. The set of solutions of the corresponding Schr\"odinger equation is also constructed and some of their basic properties are discussed.

\section{The BSS approach}
\label{sec:sam}

In this section we briefly summarize the approach proposed by Bagrov, Samsonov and Shekoyan (BSS) to generate time-dependent potentials by a Darboux transformation \cite{Bag95,Bag96}. Consider the following pair of dimensionless Schr\"odinger equations 
\begin{equation}
\left[ i \partial_{t}+\partial_{x}^{2} - V_{0}(x,t) \right] \phi(x,t) = 0,
\label{eq:sam1}
\end{equation}
\begin{equation}
\left[ i \partial_{t}+\partial_{x}^{2}-V_{1}(x,t) \right] \psi(x,t) = 0,
\label{eq:sam2}
\end{equation}
where $V_{0}(x,t)$ and $V_{1}(x,t)$ are real-valued potentials that depend on time in general. In the above equations $\partial_t$ and $\partial_x$ stand for  time and position partial-derivatives respectively. It is assumed that the operator
\begin{equation}
L=\ell(t)\left[ \beta(x,t)+\partial_{x} \right],
\label{eq:sam3}
\end{equation}
with $\ell (t)$ and $\beta(x,t)$ functions to be determined, intertwines \eqref{eq:sam1} and \eqref{eq:sam2} as follows
\begin{equation}
L  \left[ i \partial_{t}+\partial_{x}^{2}-V_{0}(x,t) \right] = \left[ i \partial_{t}+\partial_{x}^{2}-V_{1}(x,t) \right]   L.
\label{eq:sam4}
\end{equation}
Then one arrives at the set of equations
\begin{equation}
V_{1}(x,t)-V_{0}(x,t)=i\frac{\dot \ell}{\ell}+2\beta_{x}, 
\label{eq:sam5a}
\end{equation} 
\begin{equation}
 \left( \beta^{2} \right)_{x}=i\dot \beta + \beta_{xx}+(V_{0})_{x},
\label{eq:sam5b}
\end{equation} 
where $\dot f := \partial _t f$ and $f_x := \partial_x f$. Let $u$ be the transformation function, i.e. $\beta=-(\ln u)_{x}$. Therefore, (\ref{eq:sam5b}) is simplified to the Schr\"odinger-like equation
\begin{equation}
i\dot{u}=-u_{xx}+ [V_{0} + c_{1}(t) ]u,
\label{eq:sam6}
\end{equation}
with $c(t)$ an integration constant which can be set to zero. On the other hand, as the potentials $V_0$ and $V_1$ are real-valued, from (\ref{eq:sam5a}) we have $i(\ln \vert \ell(t) \vert^{2} )_{t}=-2\partial_{x}(\beta_{x}-\beta^{*}_{x})_{x}$. That is, the transformation function must satisfy
\begin{equation}
\left( \ln \frac{u}{u^{*}} \right)_{xxx} = 0.
\label{eq:sam7}
\end{equation}
Now, assuming that the function $\ell(t)$ is a real-valued we have
\begin{equation}
\ell(t)=\exp\left(2\int dt \, \textnormal{Im}(\ln \, u(x,t))_{xx} \right ), \qquad V_1(x,t) -V_0(x,t) =-\left( \ln \vert u(x,t) \vert^2 \right)_{xx}.
\label{eq:sam8}
\end{equation}
As usual in supersymmetry, the solutions $\psi(x,t)$ of \eqref{eq:sam2} can be obtained from \eqref{eq:sam4} whenever the solutions of (\ref{eq:sam1}) are given, and vice versa. Additionally, the missing state $\psi_{0}=\frac{1}{\ell u^{*}}$ must be considered since it is also a solution of \eqref{eq:sam2}.

\section{Time-dependent oscillators}
\label{sec:TDDO}

Let $V_{0}(x,t)=x^2$ be the initial potential and
\begin{equation}
u(x,t)=B(t) e^{a(t) x^2} f(x,t)
\label{boris}
\end{equation}
the transformation function, with $a(t), B(t)$ and $f(x,t)$ functions to be determined. The form of $u$ in (\ref{boris}) is a generalization of the one used in \cite{Bag95} for the free-particle potential. In our case, the $u$-function could even be adapted to be one of the Gaussian wave packets discussed in e.g. \cite{Cas13,Cru15,Cru16}. At this stage, it is convenient to introduce an additional function $z(x,t)=b(t)x$ that couples the spatial variable $x$ with the time-parameter $t$ via the real-valued function $b(t)$. The introduction of $u(z (x,t))$ into \eqref{eq:sam1} gives rise to a system which includes a differential equation in the variable $z$ for $f$, a differential equation in the variable $t$ for $B$, and a pair of constraints that are useful to determine the time-dependent functions $a$ and $b$ \cite{Zel17}. The straightforward calculation yields
\begin{equation}
f(x,t)=e^{-\frac{b^2(t) x^2}{2}} \left[ k_a \, {}_{1}F_{1}\left( \nu,\frac{1}{2};b^2(t) x^2\right)+ k_b \, b(t)x \, {}_{1}F_{1}\left(\nu+\frac{1}{2},\frac{3}{2}; b^2(t) x^2 \right) \right],
\end{equation}
with $k_a$ and $k_b$ arbitrary constants,
\begin{equation}
b(t)=\frac{c_0}{\sqrt{c_1+\gamma \cos(4t+c_2)}},
\label{bdt}
\end{equation}
and
\begin{equation}
\begin{aligned}
& B(t)=\frac{\exp\left[i\left(\nu-\frac{1}{4}\right) \frac{c_0^2}{\gamma \lambda \sqrt{1+ \lambda^2}} \arctan \left( \frac{\lambda}{\sqrt{1+ \lambda^2}}\tan\left(\frac{4t+c_2}{2} \right) \right) \right]}{\left[ c_1+ \gamma \cos(4t+c_2) \right]^{1/4}},\\[1ex]
& a(t)=-i\frac{\gamma}{2}\left(\frac{\sin(4t+c_{2})}{c_1+\gamma \cos(4t+c_2)}\right).
\end{aligned}
 \end{equation}
In the above expressions ${}_1F_1(\nu, \eta; w )$ stands for the confluent hypergeometric function in the variable $w$ \cite{Olv10} and 
 \[
\gamma^2=c_1^2-c_0^4 \, , \quad \lambda^2=\frac{1}{2}\left( c_1-1 \right),
\]
with $c_0$, $c_1$ and $c_2$ integration (real) constants such that $c_1>c_0^2$. Therefore
\[
\ell(t)=\sqrt{c_1+\gamma \cos(4t+c_2)},
\]
and the new potential is given by the time-dependent function
\begin{equation}
V_1(x,t) = x^2 + 2b^2(t) - 2 \frac{\partial^2}{\partial x^2} \left[ f(x,t) e^{\frac{b^2(t) x^2}{2}} \right].
\label{newpot}
\end{equation}
On the other hand, the well known stationary solutions of \eqref{eq:sam1},
\begin{equation}
\phi_{n}(x,t)=\frac{e^{-\frac{x^2}{2}-i(2n+1)t}}{\sqrt{2^{n}n!\sqrt{\pi}}}H_{n}(x),
\label{stationary}
\end{equation}
lead to the solutions of \eqref{eq:sam2} for the new time-dependent potential
\begin{equation}
\begin{aligned}
& \psi_{n}(x,t)=L \phi_{n+1}(x,t), \quad n=0,1,2, \ldots,\\
& \psi_{0}(x,t)=\frac{1}{\ell(t)u^{*}(x,t)}.
\end{aligned}
\label{solution}
\end{equation}

\begin{figure}[htb]

\centering
\subfigure[]{\includegraphics[width=0.35\textwidth]{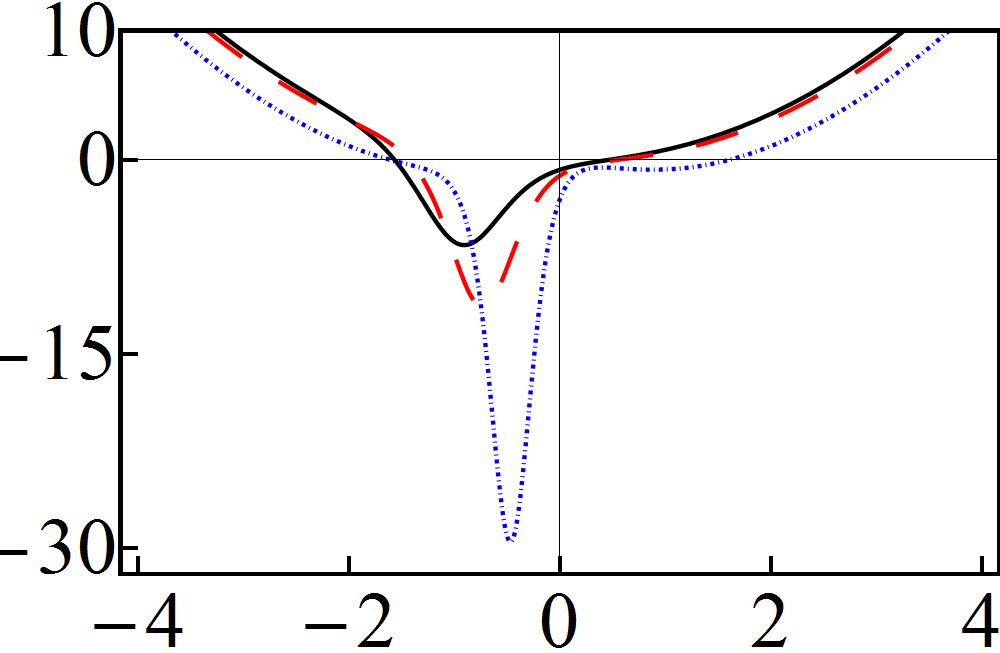}} 
\hspace{2ex}
\subfigure[]{\includegraphics[width=0.35\textwidth]{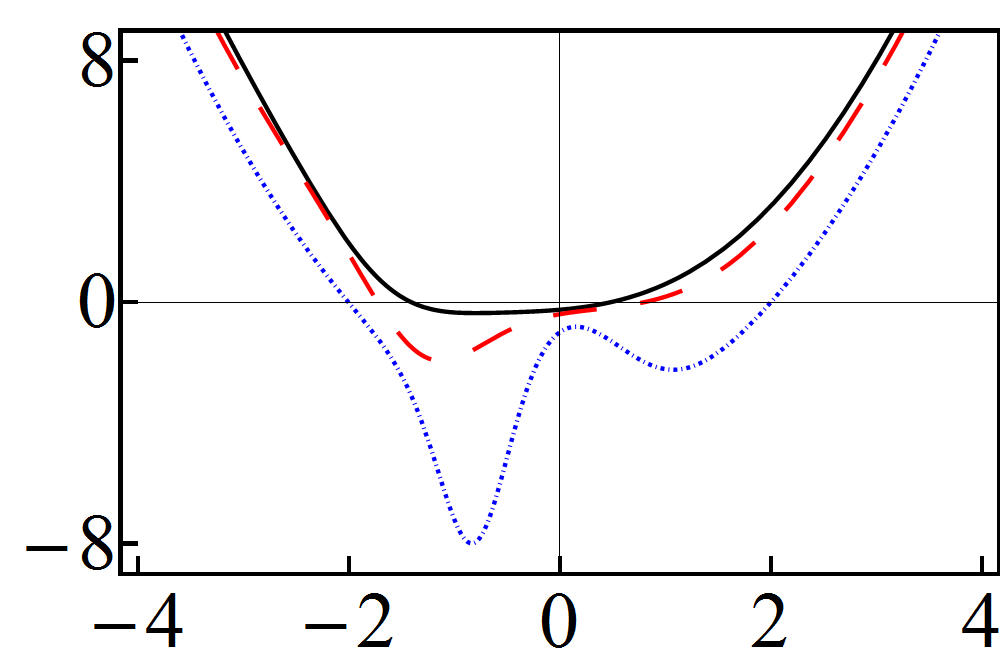}} 

\caption{\footnotesize 
(Color online)  The time-dependent oscillator potentials (\ref{newpot}) for $c_0=1$, $c_1=10$, $c_2=0$ and (a) $k_a=2$, $k_b=5$, $\nu=2$ (b) $k_a=1.3 \sqrt{\pi}$, $k_b=2$ and $\nu=1/2$. The analytic expression for the potentials depicted in (b) is given in Eq.~(\ref{mielnik}). In both cases the potential has been evaluated at $t=0$ (solid-black), $t=\frac{\pi}{8}$ (dashed-red) and  $t=\frac{\pi}{4}$ (dotted-blue).
}
\label{figpot}
\end{figure}


\subsection{Examples}

If $\nu=1/2$, the time-dependent potentials (\ref{newpot}) acquire the special form
\begin{equation}
V_{1}(x,t)=x^2- 2 b(t)-4 k_b  b(t)\frac{\partial}{\partial x}\left[ \frac{e^{-b^2(t) x^2}}{2 k_a + \sqrt{\pi}k_b \mbox{Erf} \left( b(t)x \right)} \right],
\label{mielnik}
\end{equation}
where $2k_a>\sqrt{\pi} k_b$ to avoid singularities. For $\gamma=0$ the above function reproduces  the results already reported in \cite{Mie84} for the conventional (time-independent) supersymmetric approach. 

The time-dependent potentials (\ref{newpot}) and (\ref{mielnik}) are depicted at three different times in Fig.~\ref{figpot} for the indicated sets of parameters. In both cases we can appreciate a very localized time-dependent deformation that increases in size and is shifted to the right as the time goes pass. On the other hand, the coupling function $z(x,t)=b(t)x$ is regulated by the time-dependent function defined in (\ref{bdt}). The latter oscillates between the values $b_{\pm} = c_0/\sqrt{c_1 \pm \gamma}$, so that $z$ oscillates with the time at each value of $x$. The closer is $x$ to the position of the minimum of the potential at $t=0$, the larger is the oscillation over the time.

\begin{figure}[htb]

\centering
\subfigure[]{\includegraphics[width=0.3\textwidth]{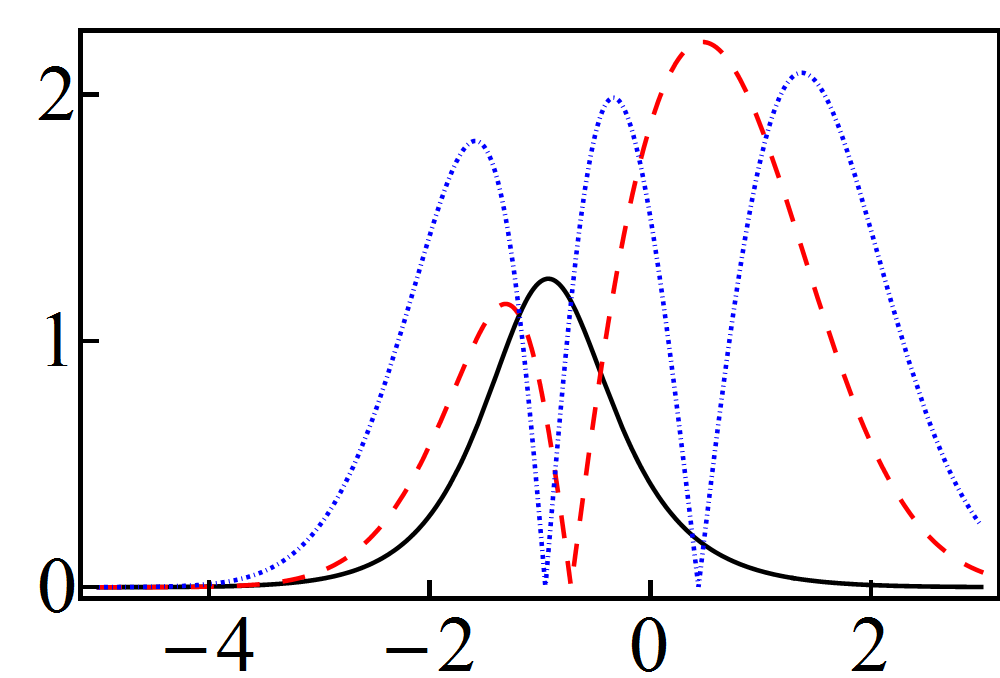}} 
\hspace{1ex}
\subfigure[]{\includegraphics[width=0.3\textwidth]{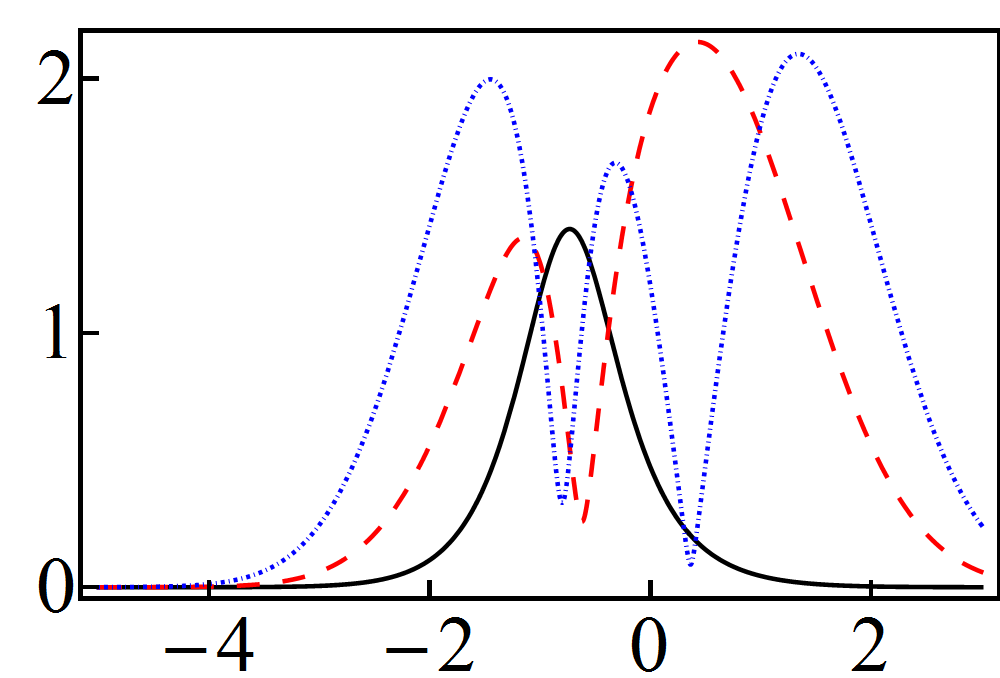}} 
\hspace{1ex}
\subfigure[]{\includegraphics[width=0.3\textwidth]{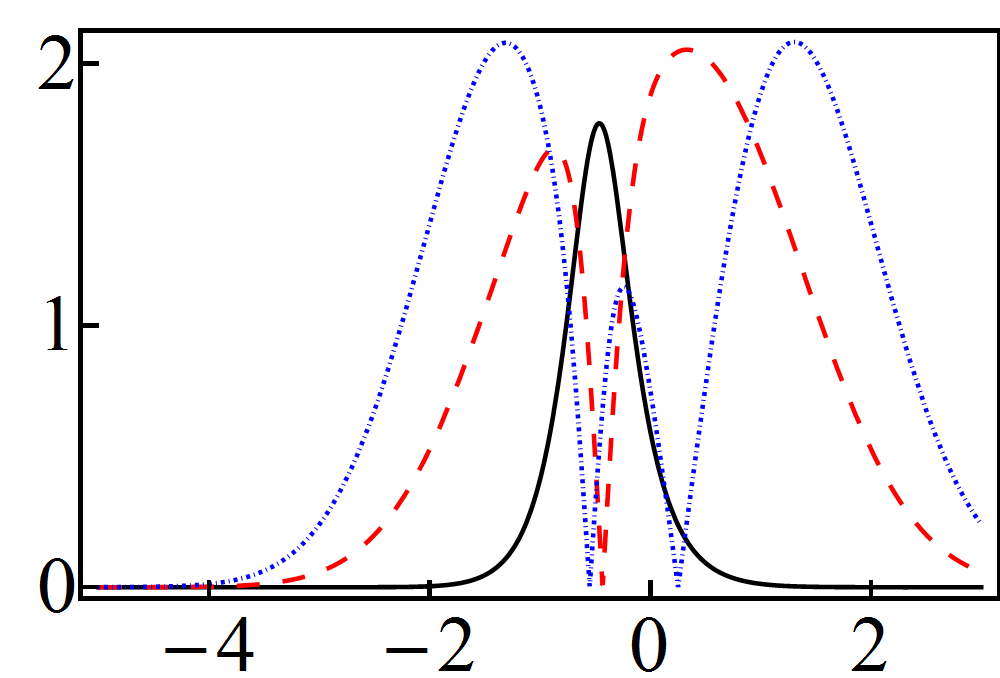}}

\centering
\subfigure[]{\includegraphics[width=0.3\textwidth]{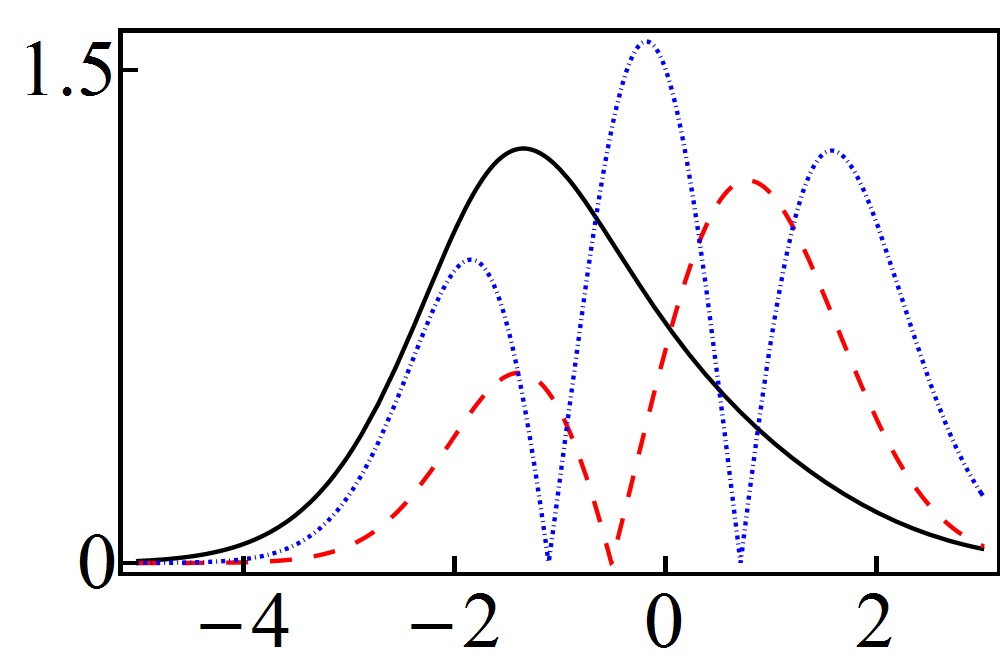}} 
\hspace{1ex}
\subfigure[]{\includegraphics[width=0.3\textwidth]{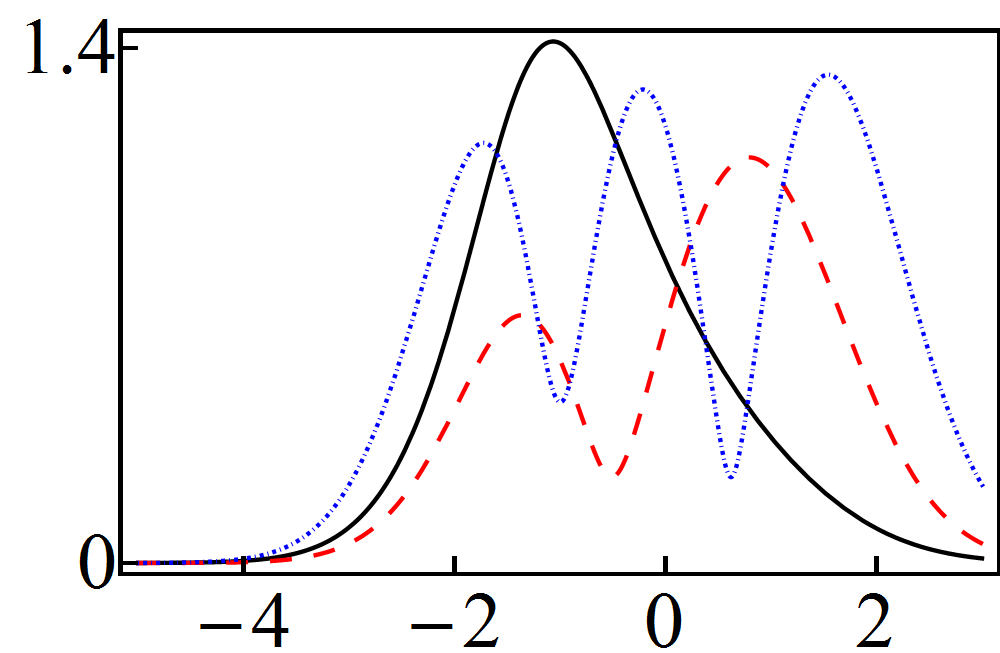}} 
\hspace{1ex}
\subfigure[]{\includegraphics[width=0.3\textwidth]{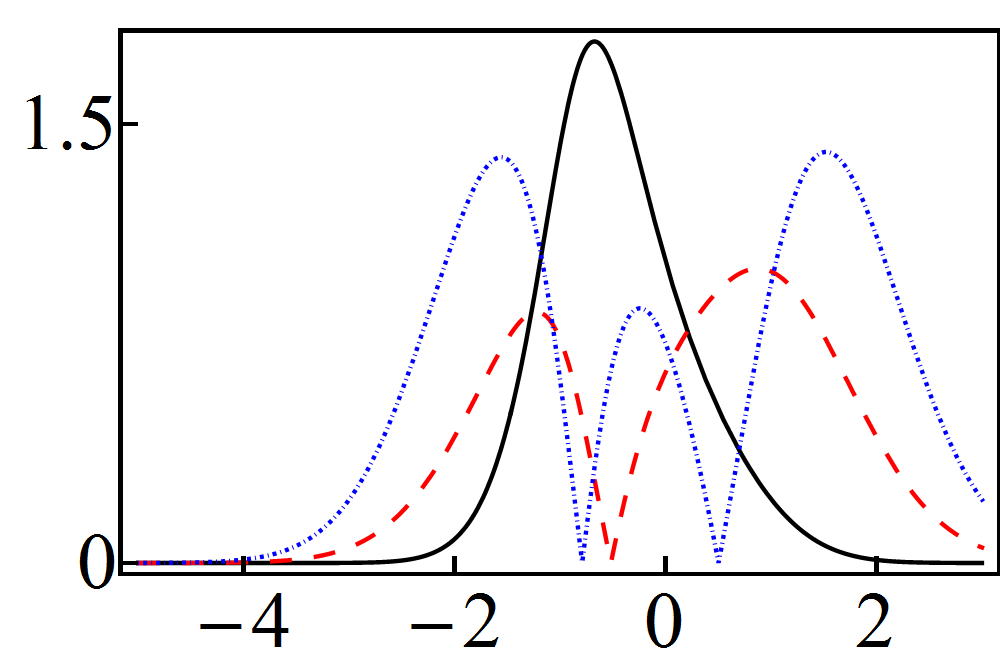}}

\caption{\footnotesize 
(Color online) Squared modulus of the non-normalized functions $\psi_0$ (solid-black), $\psi_1$ (dashed-red) and $\psi_2$ (dotted-blue), associated with the time-dependent oscillator potentials (\ref{newpot}). The upper row corresponds to the potentials depicted in Fig.~\ref{figpot}(a) and the lower row to those depicted in Fig.~\ref{figpot}(b). From left to right the columns correspond to $t=0$, $t=\frac{\pi}{8}$ and $t=\frac{\pi}{4}$.
}
\label{figsol}
\end{figure}

The squared modulus of the first three non-normalized functions (\ref{solution}) of the potentials (\ref{newpot}) and (\ref{mielnik}) are shown in Fig.~\ref{figsol} for the same parameters and times as in Fig.~\ref{figpot}. At $t=0$ and $t=\frac{\pi}4$ all of them show a distribution of zeros which obeys the well known oscillation theorem of the stationary solutions (\ref{stationary}). At other times like $t=\frac{\pi}8$, no function (\ref{solution}) has zeros. However, the distribution of the maxima of such functions follows the oscillation-like theorems obeyed by the wave-functions of complex-valued potentials \cite{Jai17}. This odd behavior deserves attention and will be analyzed elsewhere.

\section{Concluding Remarks}
\label{conc}

We have constructed time-dependent potentials which are exactly solvable and have the oscillator profile. Such potentials are Darboux-deformations of the harmonic oscillator one in the approach introduced by Bagrov, Samsonov and Shekoyan \cite{Bag95,Bag96}. We have shown that the family of oscillator-like potentials introduced by Mielnik \cite{Mie84} are included as a particular case of our results. In general, the oscillator-like potentials reported here are such that the deformation generated by the Darboux procedure oscillates with the time. We have found that the related solutions present an odd behavior since they lost their zeros in time by time. However, the maxima of these functions obey the oscillation theorems that are satisfied by the solutions to the Schr\"odinger  equation of complex-valued potentials reported in \cite{Jai17}. The progress on the latter subject and additional results will be reported elsewhere.

\section*{Acknowledgment}
K.Z. acknowledges the funding received through a CONACyT scholarship.


\end{document}